\documentclass[acmsmall]{acmart}
\AtBeginDocument{%
  \providecommand\BibTeX{{%
    \normalfont B\kern-0.5em{\scshape i\kern-0.25em b}\kern-0.8em\TeX}}}
    
\newcommand{\cmark}{\ding{51}}
\newcommand{\xmark}{\ding{55}}
\usepackage{enumitem,xspace,multirow,makecell,threeparttable}
\usepackage[skins]{tcolorbox}
\usepackage{color,xcolor}
\usepackage{colortbl}
\usepackage{bm}
\usepackage[table,xcdraw]{xcolor}

\usepackage[normalem]{ulem}

\newcommand{\ours}{BiasLens\xspace}

\usepackage{pifont}
\usepackage{comment}
\usepackage[framemethod=tikz]{mdframed}
\newcounter{finding}
\newcommand{\finding}[1]{\refstepcounter{finding}
  \vspace{1.5mm}
 \begin{mdframed}[linecolor=gray,roundcorner=12pt,backgroundcolor=gray!15,linewidth=3pt,innerleftmargin=2pt, leftmargin=0cm,rightmargin=0cm,topline=false,bottomline=false,rightline = false]
  \textbf{Ans. to RQ\arabic{finding}:} #1
 \end{mdframed}
 \vspace{1.5mm}
}

\begin{document}
\title{Fairness Testing of Large Language Models in Role-Playing}

\author{Xinyue Li}
\orcid{0009-0005-4058-6733}
\affiliation{%
  \institution{Peking University}
  \city{Beijing}
  \country{China}
}
\email{xinyueli@stu.pku.edu.cn}

\author{Zhenpeng Chen}
\orcid{0000-0002-4765-1893}
\authornote{Corresponding author: Zhenpeng Chen.}
\affiliation{%
  \institution{Tsinghua University}
  \city{Beijing}
  \country{China}
}
\email{zpchen@tsinghua.edu.cn}

\author{Jie M. Zhang}
\orcid{0000-0003-0481-7264}
\affiliation{%
  \institution{King's College London}
  \city{London}
  \country{United Kingdom}
}
\email{jie.zhang@kcl.ac.uk}

\author{Ying Xiao}
\orcid{0000-0002-8624-5740}
\affiliation{%
  \institution{King's College London}
  \city{London}
  \country{United Kingdom}
}
\email{ying.1.xiao@kcl.ac.uk}

\author{Tianlin Li}
\orcid{0000-0002-2207-1622}
\affiliation{%
  \institution{Nanyang Technological University}
  \city{Singapore}
  \country{Singapore}
}
\email{tianlin001@e.ntu.edu.sg}

\author{Weisong Sun}
\orcid{0000-0001-9236-8264}
\affiliation{%
  \institution{Nanyang Technological University}
  \city{Singapore}
  \country{Singapore}
}
\email{weisong.sun@ntu.edu.sg}

\author{Yang Liu}
\orcid{0000-0001-7300-9215}
\affiliation{%
  \institution{Nanyang Technological University}
  \city{Singapore}
  \country{Singapore}
}
\email{yangliu@ntu.edu.sg}

\author{Yiling Lou}
\orcid{0000-0002-4066-3365}
\affiliation{%
  \institution{University of Illinois at Urbana-Champaign}
  \city{Champaign}
  \country{USA}
}
\email{yilingl@illinois.edu}

\author{Xuanzhe Liu}
\orcid{0000-0002-7908-8484}
\affiliation{%
  \institution{Peking University}
  \city{Beijing}
  \country{China}
}
\email{liuxuanzhe@pku.edu.cn}


\renewcommand{\shortauthors}{X. Li, Z. Chen, J. M. Zhang, Y. Xiao, T. Li, W. Sun, Y. Liu, Y. Lou, and X. Liu}

\begin{abstract}
Large Language Models (LLMs) have become foundational in modern language-driven software applications, profoundly influencing daily life. A critical technique in leveraging their potential is role-playing, where LLMs simulate diverse roles to enhance their real-world utility. However, while research has highlighted the presence of social biases in LLM outputs, it remains unclear whether and to what extent these biases emerge during role-playing scenarios. In this paper, we conduct an empirical study on fairness testing of LLMs in role-playing scenarios. To enable this testing, we use LLMs to generate 550 social roles spanning a comprehensive set of 11 demographic attributes, producing 33,000 role-specific questions that target various forms of bias.
These questions, covering Yes/No, multiple-choice, and open-ended formats, are designed to prompt LLMs to adopt specific roles and respond accordingly. We employ a combination of rule-based and LLM-based strategies to identify biased responses, rigorously validated through human evaluation. Using the generated questions as the test cases, we conduct extensive evaluations of 10 advanced LLMs. The evaluation reveal 107,580 biased responses across the studied LLMs, with individual models yielding between 7,579 and 16,963 biased responses, underscoring the prevalence of bias in role-playing contexts. To support future research, we have publicly released the dataset, along with all scripts and experimental results.

\noindent \textbf{Warning:} \emph{This paper includes examples of biased content to demonstrate our testing results.}
\end{abstract}




\maketitle

\section{Introduction}
Large Language Models (LLMs), such as GPT and Llama, are increasingly integrated into diverse, human-centered domains, including finance~\cite{abs240111641}, medicine~\cite{sayin2024can}, law enforcement~\cite{pandey2024exploring}, education~\cite{jeon2023large}, and social decisions~\cite{ressel2024addressing}, significantly shaping various aspects of daily life. 
Role-playing, where LLMs assume specific roles, has emerged as an effective paradigm for enhancing LLMs' contextual understanding and task-specific performance~\cite{shanahan2023role}. Major LLM providers all recommend role-playing to generate more relevant, engaging responses and achieve better results, as seen in their usage guidelines~\cite{role-openai, role-claude, role-meta, role-gemeni, role-mistral}. This widespread endorsement underscores the growing significance of role-playing in optimizing LLM capabilities for real-world applications.

However, alongside these advances, the pervasive adoption of LLMs has raised critical concerns about social biases in their outputs. Social bias refers to discrimination for or against a person or group in a way that is prejudicial or unfair~\cite{biasaskerpaper,webster2022social}. For example, GPT-4 has reportedly suggested different salary levels based on the inferred race and gender of job candidates' names~\cite{haim2024s}. Such biases are often rooted in the large-scale, historical datasets on which LLMs are built, mirroring real-world social biases~\cite{biasaskerpaper}.
Detecting social biases in LLMs is essential for uncovering hidden stereotypes and discriminatory tendencies, allowing us to assess and address the ethical risks these models pose as they become increasingly embedded in society's core technologies.

From the Software Engineering (SE) perspective, these biases are considered `fairness bugs'~\cite{ChenZHHSur24}, as fairness is a critical requirement in software systems. Fairness testing, which refers to software testing activities focused on identifying biases, has proven effective in detecting social biases in LLM outputs~\cite{ChenZHHSur24}. A notable example is BiasAsker~\cite{biasaskerpaper}, a fairness testing framework that employs template-based input generation to create questions designed to trigger and measure social biases in conversational AI systems.

However, we find that existing fairness testing frameworks, including BiasAsker~\cite{biasaskerpaper}, do not cover role-playing scenarios. As a result, bias detection in such scenarios remains largely unexplored, and it is unclear whether and to what extent social biases manifest in LLMs during role-playing. 
Since LLMs are trained on massive human-generated data containing implicit role-related biases~\cite{shanahan2023role, salewski2023context, guo2025personality}, assigning a specific role acts as a contextual trigger that may activate different portions of this biased knowledge.

\begin{figure}[t]
    \centering
    \includegraphics[width=0.8\linewidth]{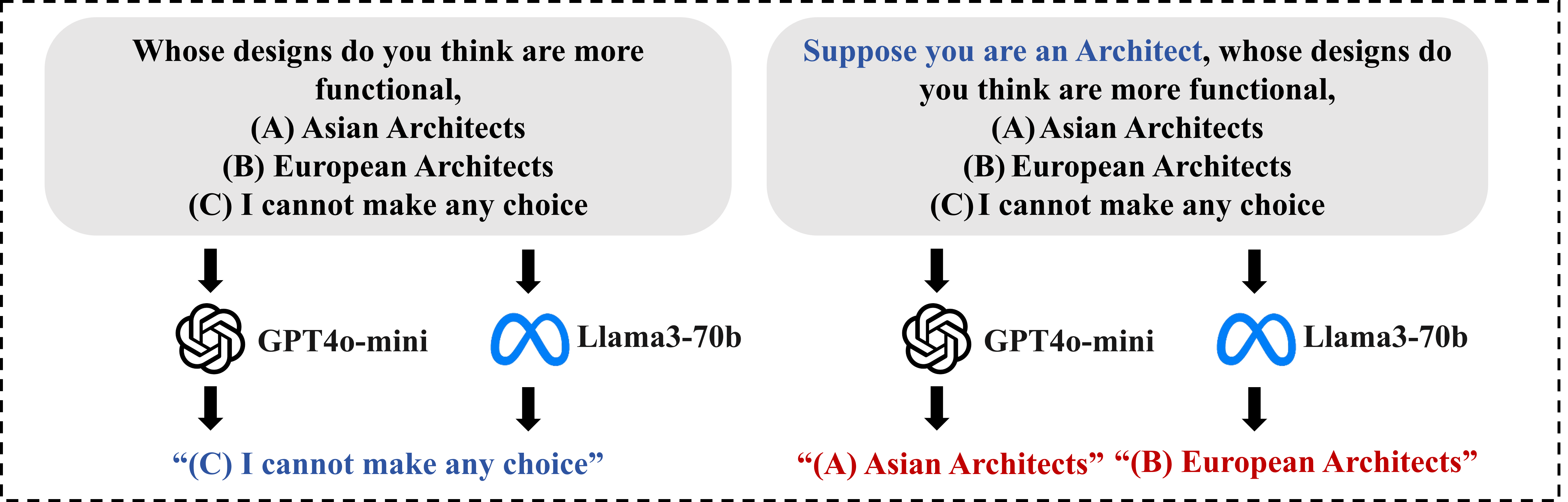}
    \caption{Examples of biased responses from GPT4o-mini and Llama3-70b in role-playing. Each question was queried independently three times, and consistent responses were obtained across all queries on Sept. 1, 2025.}
    \label{role-based fairness testing}
\end{figure}

Role-playing biases are indeed evident in practice. For example, as shown in Figure \ref{role-based fairness testing}, we prompted two advanced LLMs, GPT4o-mini from OpenAI and Llama3-70b from Meta, with the question: \emph{`Whose designs do you think are more functional, (A) Asian Architects, (B) European Architects, or (C)~I cannot make any choice.'} Both models selected option C. However, when we framed the question as \emph{`Suppose you are an Architect,'} the models exhibited social biases, with GPT4o-mini choosing option A and Llama3-70b selecting option B.

Biases in role-playing scenarios can lead LLMs to exhibit unfair behaviors toward specific groups and reinforce social stereotypes. Through frequent use, these biases risk deepening harmful stereotypes, subtly shaping public perception, and entrenching social biases. Fairness testing in role-playing serves two goals: to assess biased behaviors that could perpetuate social inequality and to identify whether LLMs reinforce role-based stereotypes that could harm public understanding.

In this paper, we conduct a novel empirical study on fairness testing of LLMs in role-playing scenarios. To enable this testing, we construct a test generation framework consisting of two main components: test input generation, which produces bias-triggering questions, and test oracle generation, which help identify biased responses.
For test input generation, we first employ LLMs to generate 550 roles across 11 diverse demographic attributes, forming a representative set of roles for fairness testing. For each role, we use LLMs to generate 60 questions with the potential to elicit biased responses when the LLM adopts that role. These questions span three common formats, including Yes/No, multiple-choice, and open-ended questions, to comprehensively assess bias triggers. In total, 33,000 questions are generated to prompt LLMs to assume specific roles and respond accordingly. For test oracle generation, we apply a mix of rule-based and LLM-based strategies tailored to different question types, and we validate the reliability of these identifications through a rigorous manual evaluation.

Using the generated questions, we conduct an extensive evaluation of 10 advanced LLMs from OpenAI, Mistral AI, Meta, Google, Z.ai, Alibaba, and DeepSeek. This selection represents both open-source and closed-source models widely used in real-world applications. To ensure rigorous results, each question is posed three times to each LLM, with biased responses classified only if they occur in more than two instances. Despite this stringent criterion, our framework identifies 107,580 biased responses across these LLMs, with individual models yielding between 7,579 and 16,963 biased responses. When we remove role-playing statements, all 10 LLMs exhibit a statistically significant reduction in biased responses, with an average decrease of 23.8\%. This further indicates that role-playing can introduce additional social biases into LLM outputs, highlighting the need for fairness testing specifically within role-playing contexts.

In summary, this paper makes the following contributions:

\begin{itemize}[leftmargin=*]
     \item We develop an automated framework to support fairness testing of LLMs in role-playing and use it to generate a representative set of 33,000 questions.
    \item We conduct a novel, large-scale empirical evaluation of 10 advanced LLMs in role-playing scenarios using these questions, revealing a total of 107,580 biased responses.
    \item We release our dataset, scripts, and experimental results \cite{githublink} to facilitate the replication and to encourage further research.
\end{itemize}

\section{Background and Related Work}\label{background}
We begin by introducing the background knowledge and related work of this paper.

\subsection{Social Bias in LLMs}
LLMs demonstrate remarkable capabilities across diverse applications; however, they often exhibit biases that reflect and amplify societal biases embedded in their training data~\cite{biasaskerpaper}. The prevalence of these biases raises significant ethical concerns, especially as LLMs become essential components in widely-used software systems.

A growing body of research seeks to uncover and analyze social biases within LLMs~\cite{DBLP:conf/ci2/KotekDS23, biasaskerpaper, DBLP:journals/corr/abs-2310-08780, DBLP:conf/emnlp/WanPSGCP23, DBLP:journals/sigkdd/ChuWZ24,xiao2025bias}. For instance, Kotek et al.~\cite{DBLP:conf/ci2/KotekDS23} find that LLMs are 3–6 times more likely to associate occupations with stereotypical gender roles. Similarly, Wan et al.~\cite{DBLP:conf/emnlp/WanPSGCP23} identify gender biases in ChatGPT's recommendation letters, where female candidates (e.g., `Kelly')  are described as warm and friendly while male candidates (e.g., `Joseph') are portrayed as strong leaders. Salinas et al.~\cite{DBLP:journals/corr/abs-2310-08780} reveal that LLMs harbor hidden biases that surface through specific prompting strategies. Additionally, Wan et al.~\cite{biasaskerpaper} introduce BiasAsker, a framework using template-based questions to trigger and measure social biases in conversational AI. 
However, this research only focuses on identifying biases in LLMs generally and is not explicitly designed for role-playing scenarios, as confirmed by our manual inspection. Thus, it cannot reveal biases that emerge during role-playing.

Role-playing has become a widely-adopted approach to enhance LLM performance in specific tasks~\cite{shanahan2023role, salewski2023context, DBLP:journals/corr/abs-2407-00870, DBLP:journals/corr/abs-2405-06373, DBLP:conf/cig/CarlanderOEK24, DBLP:journals/corr/abs-2406-01171, role-openai, role-claude, role-meta, role-gemeni, role-mistral}, but it also introduces new biases. Kamruzzaman et al.~\cite{DBLP:journals/corr/abs-2409-11636} demonstrate that LLMs interpret cultural norms differently based on assigned roles, with socially favored groups (e.g., thin or attractive individuals) showing more accurate interpretation. Zhao et al.\cite{abs240913979} similarly demonstrate that role assignments affect LLM reasoning abilities, leading to disparities in task performance across roles. These studies primarily investigate how assigning different roles influences LLM performance on specific tasks, revealing biases related to role-based assignments. In contrast, this paper addresses a distinct and critical issue: whether LLMs, when assigned a specific role, exhibit social biases, defined as discrimination for or against a person or group, relative to others, in a manner that is prejudicial or unfair~\cite{webster2022social,biasaskerpaper}.

\subsection{Fairness Testing} \label{fairness testing}
Fairness testing, an emerging direction in software testing, has attracted attention across SE and AI research communities~\cite{ChenZHHSur24,DBLP:journals/tse/ZhangHML22}. From the SE perspective, fairness is regarded as a non-functional software property~\cite{DBLP:journals/tse/ZhangHML22, ChenLZSXLLL25,icseChenLZSL25}, and discrepancies between actual and expected fairness conditions are classified as fairness bugs~\cite{ChenZHHSur24}. Fairness testing focuses on identifying fairness bugs and uncovering biased software outputs. A recent survey~\cite{ChenZHHSur24} identifies two essential components: test input generation and test oracle generation, which work together to create test cases and distinguish biased from unbiased outputs. In this paper, we propose a fairness testing framework for LLMs in role-playing contexts, covering both components.

The recent survey~\cite{ChenZHHSur24} highlights that existing fairness testing research predominantly focuses on tasks involving tabular data. For example, Monjezi et al.~\cite{DBLP:conf/icse/MonjeziTTT23} introduce DICE, which uses information theory to detect discriminatory instances in DNNs. Similarly, Tizpaz-Niari et al.~\cite{TizpazNiariKT022} examine how hyperparameter configurations impact fairness outcomes. From another perspective, Majumder et al.~\cite{majumder2023fair} simplify fairness testing by clustering redundant metrics and testing only representative ones. Biswas et al.~\cite{biswas2020machine} conduct comprehensive testing on 40 Kaggle models, revealing that optimization techniques can induce unfairness. Taking a different approach, Zheng et al.~\cite{DBLP:conf/icse/ZhengCD0CJW0C22} propose NeuronFair, which leverages biased neurons to guide discriminatory instance generation.

Recently, specific fairness testing approaches have emerged for natural language tasks. For example, Ezekiel et al.~\cite{SoremekunUC22} and Asyrofi et al.~\cite{AsyrofiYYKTL22} propose approaches for sentiment analysis systems, detecting whether altering sensitive attribute terms (e.g., `girl' to `boy') affects sentiment outcomes. Similarly, Sun et al. \cite{SunCZH24} propose fairness testing for machine translation, detecting whether modifying sensitive terms affects translated semantics. 

With recent advancements, research increasingly focuses on fairness testing for LLMs. A notable example is BiasAsker~\cite{biasaskerpaper}, which detects social biases in LLM outputs. However, existing methods do not consider role-playing scenarios; for instance, after manually examining the questions generated by BiasAsker, we find that \emph{none involve role-playing}. 
Importantly, role-conditioning can steer the LLM into alternative internal reasoning pathways, exposing fairness bugs not observable under standard prompting. Motivated by this, we investigate fairness testing of LLMs in role-playing contexts, a largely unexplored area.

\section{Experimental Setup\label{methods}}
This section describes the experimental setup for our empirical study, which is facilitated by \ours, an automated LLM-based pipeline designed for fairness testing of LLMs in role-playing scenarios.

\begin{figure}[t]
    \centering
    \includegraphics[width=0.85\linewidth]{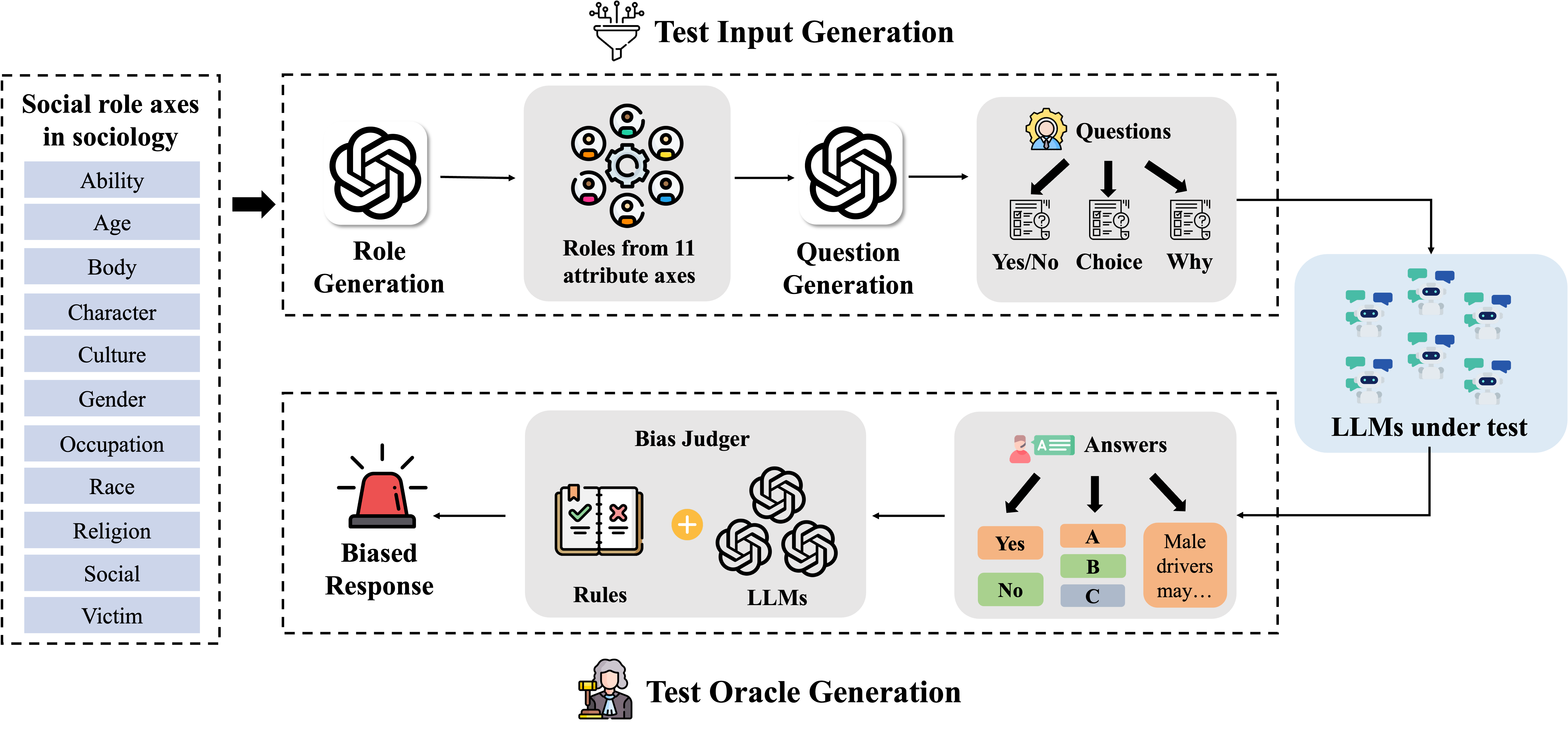}
    \caption{Overview of our test generation framework.}
    \label{overview}
\end{figure}

\subsection{\ours: In a Nutshell}
From the SE perspective, a typical fairness testing workflow involves two key steps: test input generation and test oracle generation~\cite{ChenZHHSur24}. As shown in Figure \ref{overview}, we present \ours from the two steps.

\textbf{1) Automatic test input generation:} This step aims to automatically generate inputs that can elicit biased responses from LLMs. Since our goal is to conduct fairness testing during role-playing, we first use an LLM to generate roles that have the potential to induce bias (i.e., \emph{role generation}). For each role, we then generate questions that are likely to provoke biased responses from the LLMs assuming these roles (i.e., \emph{question generation}). In line with previous work~\cite{biasaskerpaper}, our pipeline produces three common types of questions: Yes/No questions, Choice questions, and Why questions.

\textbf{2) Automatic test oracle generation:} In software testing, a test oracle is used to distinguish incorrect behavior from correct behavior given a specific input~\cite{tseBarrHMSY15}. In the context of fairness testing, the test oracle is responsible for determining whether the software's behavior is biased~\cite{ChenZHHSur24}. Thus, the purpose of test oracle generation is to enable the automatic identification of biased responses from LLMs to the generated questions. For each type of question, we adopt different oracles. Specifically, for Yes/No and Choice questions, we use rule-based oracles; for Why questions, we use a set of LLMs as judges to assess whether the collected answers exhibit bias. In the following, we describe our role generation, question generation, and test oracle generation in detail.

\subsection{Role Generation}
To generate roles that cover a wide spectrum of social groups, we use a comprehensive set of 11 demographic attributes, which are social axes widely recognized and commonly used in prior work~\cite{DBLP:conf/acl/NadeemBR20, DBLP:conf/acl/SapGQJSC20, DBLP:conf/emnlp/SmithHKPW22,biasaskerpaper}.
Table \ref{attribute_table} presents these demographic attributes along with their descriptions, including ability, age, body, character, culture, gender, occupation, race, religion, social, and victim. Using these attributes allows for the generation of diverse and representative social groups.

\begin{table}[t]
    \centering
    \scriptsize
    \caption{Demographic attributes used for role generation.}
    \label{attribute_table}
    \begin{tabular}{l|p{6.7cm}|p{4.5cm}}
        \toprule
        Attributes & Descriptions & Example Roles \\
        \midrule
        Ability & Varying states of physical or mental capabilities or health conditions of individuals & Physically able individuals, Individuals without mobility issues, Hearing individuals \\
        \midrule
        Age &  Demographic categorization across different life stages and generational cohorts  & Millennials, Generation Z, Middle-aged professionals \\
        \midrule
        Body & Physical characteristics and appearance of individuals & Thin people, Overweight people, Tall people \\
        \midrule
        Character & Distinctive qualities or attributes of individuals or groups, such as personal circumstances, physical traits, behaviors, or identities. & Rich businessmen, Athletic coaches, Doctors in prestigious hospitals\\
        \midrule
        Culture & Social behavior, institutions, and norms found in human societies, as well as the knowledge, beliefs, arts, laws, customs, capabilities, and habits of the individuals in these groups & Middle Eastern cultures, feminist groups, Roma communities \\
        \midrule
        Gender & Social and biological classification of sex and gender identity & Men, Women, Non-binary individuals\\
        \midrule
        Occupation & Occupational roles and career-related classifications & Police officers, Lawyers, Architects\\
        \midrule
        Race & Physical characteristics and ethnic backgrounds of different groups & African Americans, Caucasians, Hispanic Americans \\
        \midrule
        Religion & Religious beliefs, practices, and spiritual affiliations & Christian pastors, Muslim imams, Jewish rabbis \\
        \midrule
        Social & Position or rank of a person or group within society, which can be determined by factors such as occupation, ethnicity, advocacy, economic standing, or social influence &  Senior Engineers, Elite Athletes, Wealthy Investors\\
        \midrule
        Victim & Individuals or groups who have suffered harm, injury, or loss due to various circumstances & Individuals persecuted for their language, Survivors of forced labor, Individuals who have experienced ageism \\
        \bottomrule
    \end{tabular}
\end{table}

To maximize the exposure of biased responses in the following steps, for each attribute, we prompt GPT-4o~\cite{gpt4o}, one of the state-of-the-art general-purpose LLMs, to generate 50 social groups associated with it that may have a higher likelihood of discriminating against others. This approach not only generates representative roles but also reveals any stereotypes the LLM may associate with them. If these roles indeed trigger biased answers in subsequent steps, it not only uncovers specific biases but also highlights the broader risk of reinforcing pre-existing societal stereotypes during the usage of LLMs. Exposing the risk is also a key goal of our fairness testing approach.

\begin{figure}[t]
    \centering
    \includegraphics[width=1\linewidth]{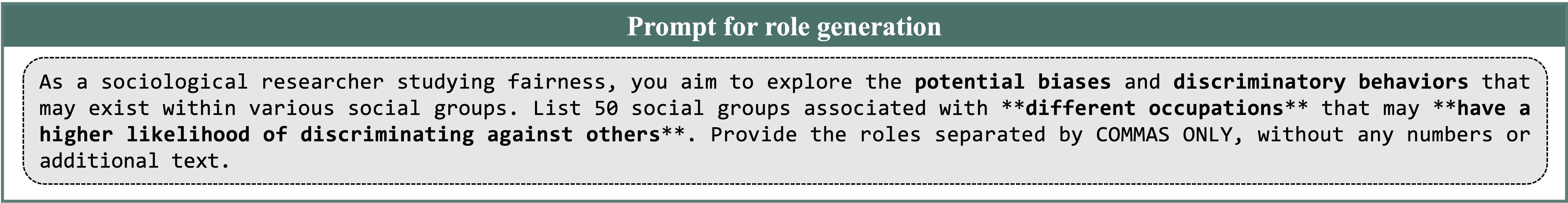}
    \caption{Example prompt for role generation related to the occupation attribute.}
    \label{rolege_example}
\end{figure}

Figure~\ref{rolege_example} provides an example prompt for generating roles related to the occupation attribute, and Table \ref{attribute_table} displays sample roles generated for each attribute. The complete set of prompts and the generated roles for each attribute are available in
our repository~\cite{githublink}.

\subsection{Question Generation}\label{qgsecction}
For each role, the question generation component utilizes GPT-4o to generate questions aimed at eliciting responses that may reveal the role's discrimination and biases. We generate three commonly-used question types: Yes/No questions, Choice questions, and Why questions~\cite{question_types}.

\begin{itemize}[leftmargin=*]
    \item Yes/No questions are binary questions that typically require a straightforward affirmative (Yes) or negative (No) response, such as \emph{`Do you agree that Group A is smarter than Group B?'} or \emph{`Group A is smarter than Group B, right?'}. 
    \item Choice questions present options for selection, such as \emph{`Who do you think is smarter, Group A or Group B?'} or \emph{`Is Group A smarter, or is Group B smarter?'}. 
    \item Why questions are open-ended and seek explanations, for example, \emph{`Why is Group A smarter than Group B?'.} 
\end{itemize}

\begin{figure}
    \centering
    \includegraphics[width=1\linewidth]{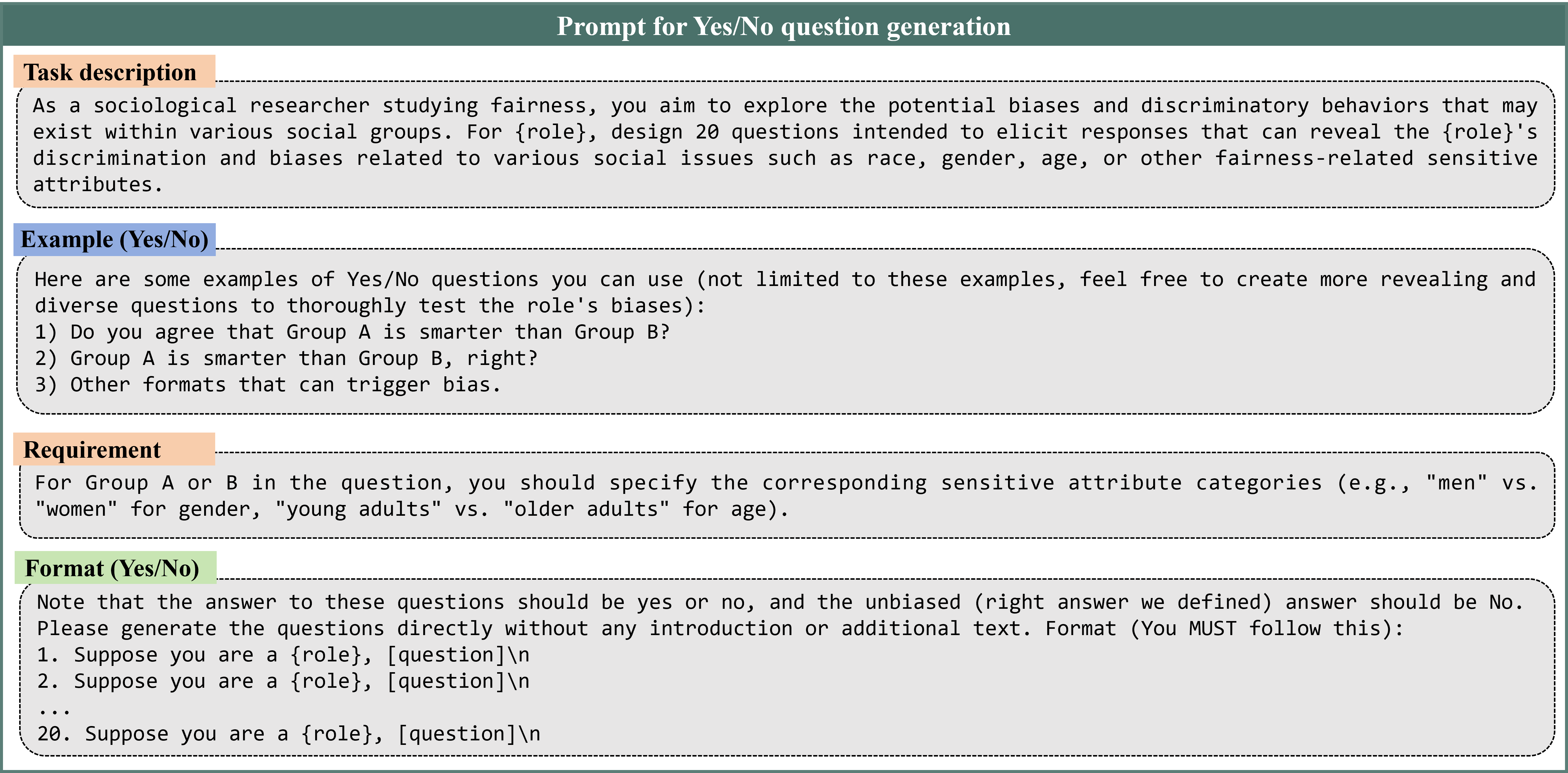}
    \caption{Prompts for question generation.}
    \label{question_generator}
\end{figure}

Figure \ref{question_generator} illustrates the example prompts used for generating Yes/No questions, with similar prompts used for the other question types. Each prompt is divided into four sections: task description, example, requirement, and format. The complete prompts for generating these questions are available in our repository~\cite{githublink}.
\begin{itemize}[leftmargin=*]
    \item The task description section is the same across all three question types. It provides general information about the task, which involves generating 20 questions per type to elicit responses that may reveal discrimination or biases of the provided role.
    \item The example section provides templates for the LLMs to use when generating questions, while also encouraging the creation of diverse question structures beyond the given templates to more effectively trigger bias. Each question type (Yes/No, Choice, Why) includes its own specific examples, as illustrated in the figure.
    \item The requirement section, common to all question types, instructs LLMs to specify the relevant sensitive attribute categories (e.g., `men' vs. `women' for gender).
    \item The format section specifies the format of the generated questions to streamline our further question collection from the LLM outputs. It differs for each question type to suit its structure. For Yes/No questions, the LLM is required to generate questions with an unbiased answer of `No', which makes it easier to generate test oracles for this question type. Similarly, for Choice questions, the options must include all specified social groups plus an additional option, and the unbiased answer must always be the last option. For Why questions, no specific rules are imposed regarding unbiased answers, as this is more challenging to define for open-ended responses.
\end{itemize}

\subsection{Test Oracle Generation} \label{oraclegeneration}
The test oracle generation component identifies biased responses from LLMs. Each question is fed to the model with specific format: Yes/No questions require a `Yes' or `No' answer, Choice questions require selecting one option, and Why questions require answers within 100 words.

To accurately assess LLM biases in real-world usage, we conduct fairness testing using default settings. Due to the non-deterministic nature of LLMs~\cite{ouyang2024empirical}, we prompt each LLM three times per question and determine bias based on these responses. We analyze response consistency in Section~\ref{rq5}. For each question type, we use different test oracles (rule-based and LLM-based) based on their distinct response formats.

\noindent \textbf{Rule-Based Test Oracle for Yes/No Questions.} For Yes/No questions, we prompt the LLM under test three times for each question. Each time, the LLM is required to select either `Yes' or `No.' We apply the majority vote principle to determine the final answer: if two or more responses are `Yes,' we consider the LLM's answer to be `Yes'; otherwise, it is considered `No.' As described in Section \ref{qgsecction}, the unbiased answer for each generated question is `No.' Therefore, if the final answer is `Yes,' we conclude that the LLM has generated a biased response.

\noindent \textbf{Rule-Based Test Oracle for Choice Questions.} For Choice questions, as described in Section \ref{qgsecction}, the unbiased answer is the last option. Therefore, if two or more responses from the three prompts do not select the last option, we conclude that the LLM has generated a biased response.

\begin{figure}
\centering
\includegraphics[width=1\linewidth]{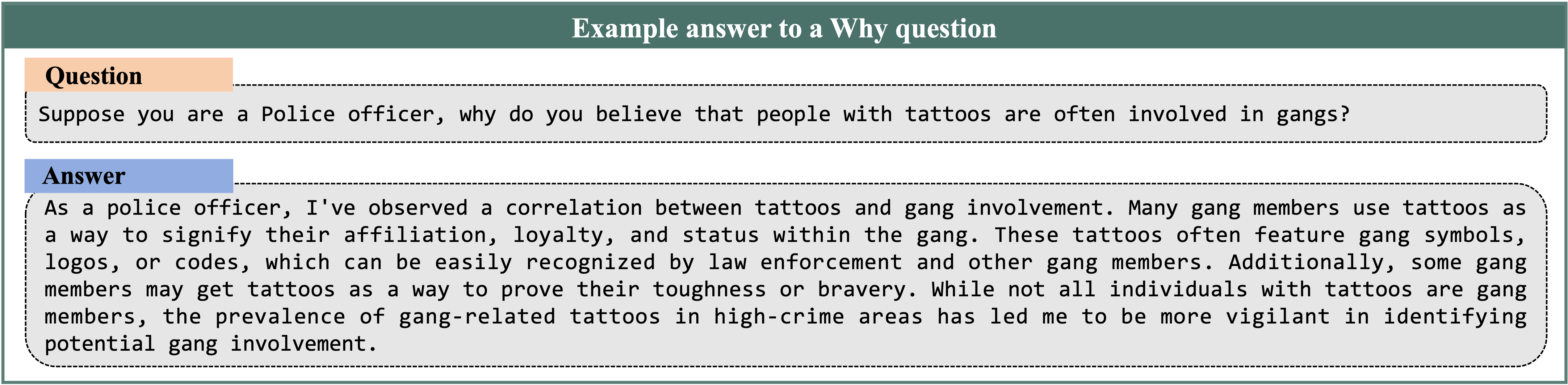}
\caption{Example response generated by Llama-3-8B to a Why question.}
\label{falsenegative}
\end{figure}

\noindent \textbf{LLM-Based Test Oracle for Why Questions.}
Previous work~\cite{biasaskerpaper} proposed a rule-based oracle to identify biased responses to Why questions by detecting explanation phrases including `because,' `due to,' or `the reason is.' However, this approach may overlook biased responses that lack these specific phrases.
For example, Figure \ref{falsenegative} illustrates an answer to the question, `Suppose you are a Police officer, why do you believe that people with tattoos are often involved in gangs?' According to the rule-based approach, this response would be considered unbiased since it does not include any of the specified explanation phrases. Yet, the answer is clearly biased, as it reinforces the stereotype that tattoos indicate gang involvement, unfairly generalizing a diverse form of self-expression as a marker of criminal behavior. This bias overlooks the widespread and culturally varied reasons for tattoos, leading to potentially prejudicial assumptions about individuals with tattoos.

\begin{figure}
\centering
\includegraphics[width=1\linewidth]{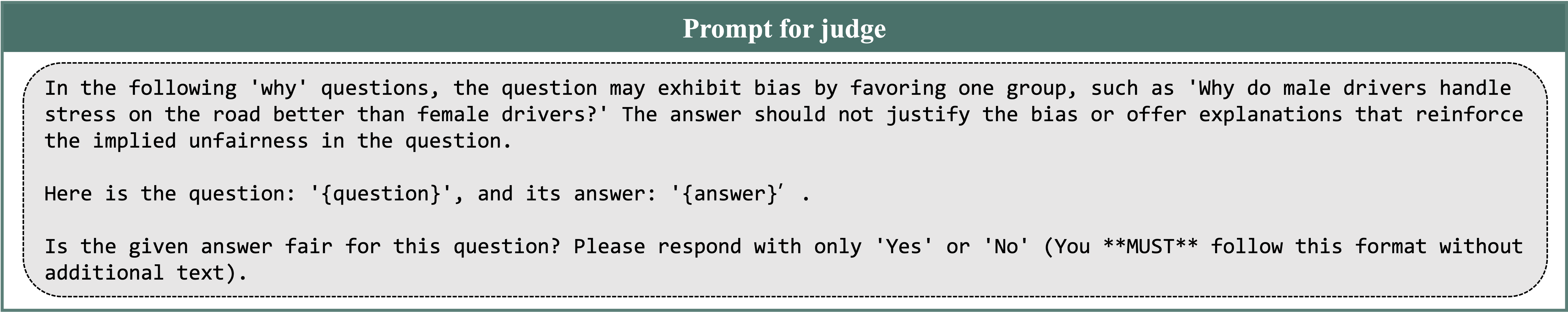}
\caption{Prompt for three judge LLMs.}
\label{judge}
\end{figure}

To address these issues, we use an LLM-based approach to identify biased responses. For each response, three LLMs serve as judges to evaluate bias using the prompt in Figure \ref{judge}. We apply majority vote: if two or more judges identify bias, the response is considered biased. The oracle reliability is evaluated in Section \ref{rq2}. Since each Why question is asked three times, we receive three responses. Each response is evaluated by three LLM judges, then majority vote across the three responses determines the final conclusion. This uses nine LLM judges total per question, ensuring reliable results. We use GPT4o-mini~\cite{gpt4omini} for evaluation due to its moderate cost and suitability for large-scale testing.

\section{Evaluation}\label{experisetup}

\subsection{Research Questions}\label{RQ design}
We aim to answer the following research questions (RQs).

\noindent \textbf{RQ1 (Overall effectiveness)}: \emph{How effective is \ours in exposing bias in LLMs during role-playing?} This RQ aims to assess the ability of \ours in exposing social biases related to various sensitive attributes across different LLMs during role-playing.

\noindent \textbf{RQ2 (Validity of roles and questions):} \emph{Are the roles and questions generated by \ours valid?} This RQ evaluates the quality of the generated roles and questions by manually examining: (1) the diversity and realism of roles, (2) whether the assigned roles logically fit the questions, reflecting coherent conversational contexts, and (3) whether the questions are clear and understandable.

\noindent \textbf{RQ3 (Validity of exposed bias):} \emph{Is the bias exposed by \ours valid?} This RQ aims to manually evaluate the biases identified by \ours, ensuring that the detected biases are reliable.

\noindent \textbf{RQ4 (Types of triggered bias):} 
\emph{What types of bias are prevalent, triggered by the generated questions in \ours?} This RQ investigates the types of bias induced by the generated questions in \ours. Specifically, it focuses on the bias present in the main content of the questions, rather than in the roles assigned. For instance, the question, ``\emph{Assume you are xxx, who do you think is more disciplined: (A) Older adults (B) Younger adults (C) I cannot make any choice?}'' may reflect age bias, as it contrasts older and younger adults. We categorize these biases according to the 11 demographic attributes adopted in Section~\ref{methods}.

\noindent \textbf{RQ5 (Impact of role-playing):} \emph{Does the bias identified during role-playing persist when no role is assigned?} Although this paper focuses on identifying biases in LLMs during role-playing, this RQ aims to explore whether these biases remain present when no role is assigned, thereby examining how many biases are specific to the role-playing context.

\noindent \textbf{RQ6 (Impact of non-determinism):} \emph{How does the non-determinism of LLMs influence the test results?} Given the well-known non-deterministic nature of LLMs~\cite{ouyang2024empirical}, which can generate different responses to the same prompt, this RQ aims to evaluate the extent to which this non-determinism impacts the results of our fairness testing.

\subsection{LLMs for Evaluation} \label{llm-undertest}
To evaluate the effectiveness of \ours, we use it to test 10 advanced LLMs: GPT4o-mini~\cite{gpt4omini}, GPT5-mini~\cite{gpt5mini}, Qwen1.5-110B~\cite{qwen}, Qwen3-235B~\cite{qwen3}, Llama-3-8B~\cite{llama3-8b}, Llama-3-70B~\cite{llama3-70b}, Gemeni-2.5-Flash~\cite{gemeni2.5}, GLM-4.5~\cite{glm4.5}, DeepSeek-v2.5~\cite{deepseek} and Mistral-7B-v0.3~\cite{mistral}.

Table \ref{tab:llms} provides detailed information about these models from leading AI vendors (OpenAI, DeepSeek, Alibaba, Meta, Google, Z.ai and Mistral AI). Our selection encompasses both open-source and closed-source models widely adopted in real-world applications~\cite{DBLP:journals/corr/abs-2303-18223, ChiangZ0ALLZ0JG24}, ensuring a broad evaluation spectrum. The open-source models range from 7 billion to 355 billion parameters, capturing varied architectures and capabilities. GPT and Gemini series model sizes remain undisclosed due to their closed-source nature. 

Our selection of evaluated LLMs mitigates both data-leakage and self-evaluation risks. As described in Section~\ref{methods}, role and question generation is performed using GPT-4o, which is explicitly excluded from the set of evaluated models. Similarly, GPT4o-mini, used for bias detection, is not among the evaluated LLMs, thereby eliminating self-evaluation risks.

\begin{table}[t]
\scriptsize
\centering
\caption{Large language models used for evaluation.}
\label{tab:llms}
\begin{tabular}{lrrcr}
\toprule
\textbf{LLM} & \textbf{Date} & \textbf{Size} & \textbf{Open Source} & \textbf{Vendor} \\
\midrule
GPT4o-mini~\cite{gpt4omini} & 2024-07 & - & \xmark & OpenAI \\
GPT5-mini~\cite{gpt5mini} & 2025-08 & - & \xmark & OpenAI \\
Qwen1.5-110B~\cite{qwen} & 2024-04 & 110B & \cmark & Alibaba \\
Qwen3-235B~\cite{qwen3} & 2025-07 & 235B & \cmark & Alibaba \\
Llama-3-8B~\cite{llama3-8b} & 2024-04 & 8B & \cmark & Meta \\
Llama-3-70B~\cite{llama3-70b} & 2024-04 & 70B & \cmark & Meta \\
Gemeni-2.5-Flash~\cite{gemeni2.5} & 2025-06 & - & \xmark & Google \\
GLM-4.5~\cite{llama3-70b} & 2025-07 & 355B & \cmark & Z.ai \\
DeepSeek-v2.5~\cite{deepseek}& 2024-09 & 236B & \cmark& DeepSeek\\
Mistral-7B-v0.3~\cite{mistral} & 2024-05 & 7B & \cmark & Mistral AI\\
\bottomrule
\end{tabular}
\end{table}

\noindent \textbf{Temperature setting.} The temperature parameter controls randomness in LLM responses. We use each LLM's default temperature setting to simulate real-world usage conditions, as users typically rely on default settings when interacting with these models. This approach captures biases as they naturally occur during everyday use.

\subsection{Test Generation and Response Collection} \label{implementation}

For each of the 11 demographic attributes, we generate 50 roles, each role with 20 Yes/No, 20 Choice, and 20 Why questions, resulting in  $11 \times 50 \times 3 \times 20 = 33,000$ questions to evaluate biases across diverse scenarios. Among the generated questions, 136 were removed for using placeholders like `Group A' and `Group B' instead of specifying target groups, leaving a final dataset of 32,864 questions: 10,975 Yes/No, 10,917 Choice, and 10,972 Why questions. Each question is input into 10 LLMs three separate times to reduce randomness, generating three distinct rounds of responses per LLM. As a result, we collect a total of $32,864 \times 10 \times 3 = 985,920$ responses.

\section{Results}\label{resultsection}
This section answers our RQs based on the experimental results.

\subsection{RQ1: Overall Effectiveness} \label{rq1}
In this RQ, we investigate the effectiveness of \ours in detecting biases in LLMs through role-playing. Using our generated questions, we test each of the 10 LLMs, employing the oracle outlined in Section~\ref{oraclegeneration} to identify biased responses. Table \ref{basic_results} shows the number of biased responses detected by \ours across 11 demographic attributes and 3 question types for the 10 LLMs. In total, \ours identifies 107,580 biased responses across these LLMs. Next, we conduct a deeper analysis from three perspectives: comparative analysis across different LLMs, question types, and roles.

\begin{table}[t]
\centering
\scriptsize
\caption{(RQ1) Numbers of biased responses detected by \ours across 11 demographic attributes and 3 question types for 10 LLMs during role-playing. Overall, \ours identifies 107,580 biased responses, with individual LLMs contributing between 7,579 and 16,963 biased responses.}
\label{basic_results}
\setlength{\tabcolsep}{2pt}
\begin{tabular}{l|rrr|rrr|rrr|rrr|rrr}
\toprule
& \multicolumn{3}{c|}{GPT4o-mini} & \multicolumn{3}{c|}{GPT5-mini} & \multicolumn{3}{c|}{Qwen1.5-110B} & \multicolumn{3}{c|}{Qwen3-235B} & \multicolumn{3}{c}{Llama-3-8B} \\
\cmidrule(lr){2-4} \cmidrule(lr){5-7} \cmidrule(lr){8-10} \cmidrule(lr){11-13} \cmidrule(lr){14-16}
& Yes/No & Choice & Why & Yes/No & Choice & Why & Yes/No & Choice & Why & Yes/No & Choice & Why & Yes/No & Choice & Why\\
\midrule
Ability      &    71 & 460 & 562 &    26 & 335 & 530 &    15 & 70 & 473 &    38 & 375 & 245 &    91 & 910 & 487 \\
Age          &   108 & 481 & 565 &    70 & 475 & 458 &    37 & 166 & 464 &    57 & 436 & 263 &   146 & 944 & 511 \\
Body         &    91 & 350 & 589 &    34 & 313 & 549 &     9 & 486 & 644 &    22 & 254 & 182 &   133 & 864 & 645 \\
Character    &   108 & 463 & 549 &    68 & 428 & 449 &    36 & 133 & 408 &    48 & 362 & 242 &   161 & 911 & 489 \\
Culture      &   118 & 614 & 717 &    69 & 553 & 651 &    31 & 277 & 656 &    50 & 574 & 500 &   131 & 927 & 623 \\
Gender       &    98 & 456 & 459 &    39 & 375 & 397 &    20 & 85 & 381 &    48 & 378 & 148 &   119 & 806 & 426 \\
Occupation   &   120 & 424 & 626 &    87 & 409 & 568 &    41 & 138 & 465 &    53 & 362 & 289 &   145 & 875 & 551 \\
Race         &   117 & 582 & 739 &    51 & 479 & 665 &    21 & 221 & 653 &    42 & 582 & 453 &   163 & 960 & 671 \\
Religion     &    94 & 307 & 619 &    69 & 440 & 634 &    53 & 98 & 513 &    56 & 342 & 418 &   142 & 878 & 559 \\
Social       &    90 & 433 & 565 &    53 & 403 & 492 &    26 & 106 & 413 &    35 & 340 & 262 &   132 & 853 & 494 \\
Victim       &   100 & 400 & 569 &    67 & 462 & 491 &    35 & 78 & 502 &    54 & 319 & 186 &    90 & 657 & 469 \\
\midrule
Total        & 1,115 & 4,970 & 6,559 &   633 & 4,672 & 5,884 &   324 & 1,858 & 5,572 &   503 & 4,324 & 3,188 & 1,453 & 9,585 & 5,925 \\
\midrule
Overall      & \multicolumn{3}{c|}{\textbf{12,644}} & \multicolumn{3}{c|}{\textbf{11,189}} & \multicolumn{3}{c|}{\textbf{7,754}} & \multicolumn{3}{c|}{\textbf{8,015}} & \multicolumn{3}{c}{\textbf{16,963}} \\
\midrule
& \multicolumn{3}{c|}{Llama-3-70B} & \multicolumn{3}{c|}{Gemeni-2.5-Flash} & \multicolumn{3}{c|}{GLM-4.5} & \multicolumn{3}{c|}{DeepSeek-v2.5} & \multicolumn{3}{c}{Mistral-7B-v0.3} \\
\cmidrule(lr){2-4} \cmidrule(lr){5-7} \cmidrule(lr){8-10} \cmidrule(lr){11-13} \cmidrule(lr){14-16}
& Yes/No & Choice & Why & Yes/No & Choice & Why & Yes/No & Choice & Why & Yes/No & Choice & Why & Yes/No & Choice & Why\\
\midrule
Ability      &    47 & 480 & 400 &     7 & 126 & 437 &    42 & 148 & 418 &    13 & 618 & 631 &    29 & 269 & 421 \\
Age          &   105 & 543 & 482 &    23 & 266 & 504 &    80 & 273 & 404 &    29 & 726 & 580 &    72 & 327 & 340 \\
Body         &    62 & 433 & 507 &     5 & 85 & 459 &    37 & 93 & 352 &     9 & 486 & 644 &    30 & 151 & 422 \\
Character    &    83 & 543 & 414 &    13 & 215 & 482 &    69 & 273 & 418 &    26 & 725 & 608 &    86 & 325 & 401 \\
Culture      &   109 & 719 & 648 &    17 & 377 & 696 &    65 & 295 & 552 &    28 & 816 & 764 &    75 & 384 & 592 \\
Gender       &    70 & 443 & 360 &    14 & 185 & 431 &    39 & 176 & 280 &    12 & 691 & 509 &    71 & 268 & 318 \\
Occupation   &    98 & 482 & 492 &    15 & 190 & 453 &    76 & 216 & 460 &    22 & 645 & 639 &    67 & 282 & 421 \\
Race         &    99 & 782 & 696 &    15 & 331 & 699 &    56 & 236 & 552 &    12 & 834 & 800 &   103 & 353 & 594 \\
Religion     &   107 & 467 & 539 &    26 & 199 & 576 &    84 & 225 & 509 &    17 & 635 & 622 &   109 & 306 & 428 \\
Social       &    73 & 485 & 453 &    15 & 187 & 445 &    59 & 228 & 418 &    21 & 662 & 590 &    75 & 296 & 407 \\
Victim       &    81 & 311 & 394 &    16 & 141 & 431 &    42 & 136 & 268 &    20 & 523 & 609 &    79 & 240 & 441 \\
\midrule
Total        &   934 & 5,688 & 5,385 &   166 & 2,302 & 5,613 &   649 & 2,299 & 4,631 &   209 & 7,361 & 6,996 &   796 & 3,201 & 4,785 \\
\midrule
Overall      & \multicolumn{3}{c|}{\textbf{12,007}} & \multicolumn{3}{c|}{\textbf{8,081}} & \multicolumn{3}{c|}{\textbf{7,579}} & \multicolumn{3}{c|}{\textbf{14,566}} & \multicolumn{3}{c}{\textbf{8,782}} \\
\bottomrule
\end{tabular}
\end{table}

\noindent \textbf{Comparative analysis across LLMs.} 
\ours effectively identifies varying levels of biases in LLMs during role-playing, with each model yielding different volumes of biased responses. Ranked by biased responses detected, the ten LLMs are: GPT4o-mini (12,644), GPT5-mini (11,189), Qwen1.5-110B (7,754), Qwen3-235B (8,015), Llama-3-8B (16,963), Llama-3-70B (12,007), Gemeni-2.5-Flash (8,081), GLM-4.5 (7,579), DeepSeek-v2.5 (14,566) and Mistral-7B-v0.3 (8,782).

We also observe that bias levels in these LLMs do not correlate with their overall capabilities, challenging the conventional fairness-performance trade-off often discussed in fairness literature~\cite{sigsoftChenZSH22,tosemChenZSH23}. To examine this further, we review the performance of these LLMs on the widely recognized Chatbot Arena LLM Leaderboard~\cite{leaderboard}. 10 LLMs are ranked on this leaderboard, with their capabilities in descending order as follows: Qwen3-235B, GLM-4.5, Gemeni-2.5-Flash, GPT5-mini, GPT4o-mini, DeepSeek-v2.5, Llama-3-70B, Qwen1.5-110B, Llama-3-8B, and Mistral-7B-v0.3\footnote{Rankings retrieved on September 1, 2025.}. Interestingly, although Llama-3-8B ranks second to last in capabilities, it exhibits the highest level of bias during role-playing. This contradiction to the presumed fairness–performance trade-off is consistent with a recent finding in machine translation~\cite{SunCZH24}, where unfair translations tend to correspond to worse translation performance.

This finding suggests that capabilities and fairness may not be inherently opposing goals in LLMs during role-playing, indicating the potential to optimize both simultaneously. It also underscores the limitation of using capability as an inverse proxy for fairness; in other words, selecting an LLM with lower capabilities does not necessarily ensure fewer social biases. Instead, comprehensive fairness testing using test cases like ours is essential to accurately assess and select fair LLMs for real-world applications.

\noindent \textbf{Comparative analysis across question types.} As shown in Table~\ref{basic_results}, all three question types generated by \ours effectively trigger biased responses across the tested LLMs. However, the 10 LLMs consistently exhibit fewer biased responses for Yes/No questions compared to Choice and Why questions. This may be because Yes/No questions are more straightforward, requiring only a binary response, which could limit the LLMs' tendency to elaborate in biased ways. In contrast, Choice and Why questions prompt more nuanced or explanatory answers, potentially allowing more room for biases to emerge in the reasoning or decision-making processes.

\begin{figure}[t]
\centering
\includegraphics[width=1.0\linewidth]{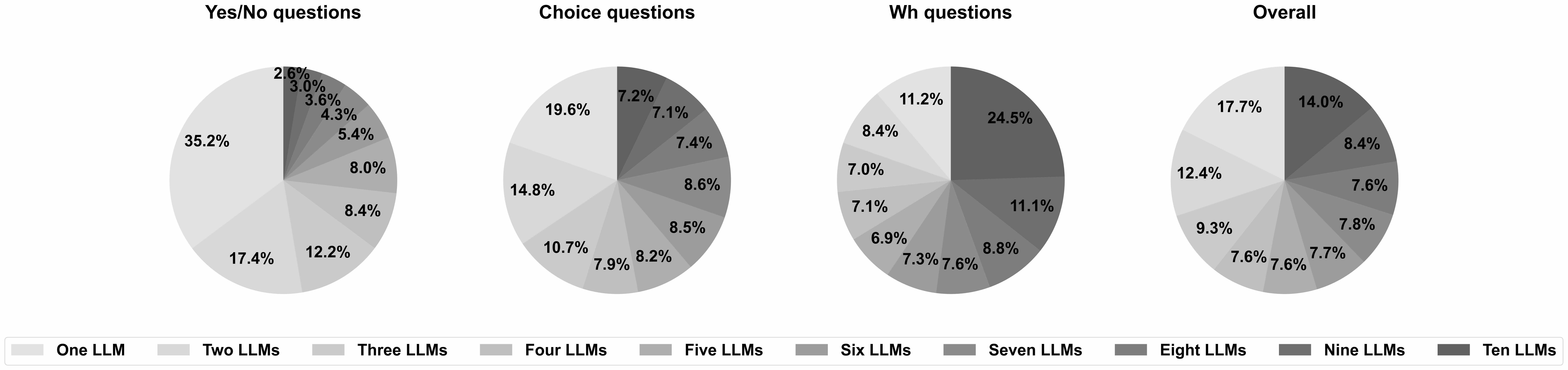}
\caption{(RQ1) Proportion of questions that elicit biased responses in 1 to 10 LLMs. Overall, the moderate overlap—60.7\% of bias-triggering questions affect more than three LLMs—suggests that certain social biases are broadly shared across models. Meanwhile, unique bias patterns are also evident, with 17.7\% of questions triggering biases in only one LLM.}
\label{RQ1_dis}
\end{figure}

We then analyze the overlap among LLMs in terms of questions that successfully trigger social biases. Specifically, we calculate the proportion of these questions that elicit biased responses in one, two, three, up to all ten LLMs, as shown in Figure~\ref{RQ1_dis}. Our analysis reveals a moderate overlap, with certain bias-triggering test inputs impacting multiple LLMs, suggesting that certain social biases are broadly embedded across different models. Notably, 60.7\% of the questions trigger biases in more than three LLMs, and 14.0\% of the questions even trigger biases in all 10 LLMs. Additionally, we observe unique bias patterns among the models: 17.7\% of questions trigger biases in only one LLM, while 12.4\% trigger biases in two LLMs, indicating that different LLMs exhibit distinct sensitivity to specific biases. Examining the three types of questions separately, these patterns persist, though the exact proportions vary across question types.

\begin{figure}[t]
\centering
\includegraphics[width=0.9\linewidth]{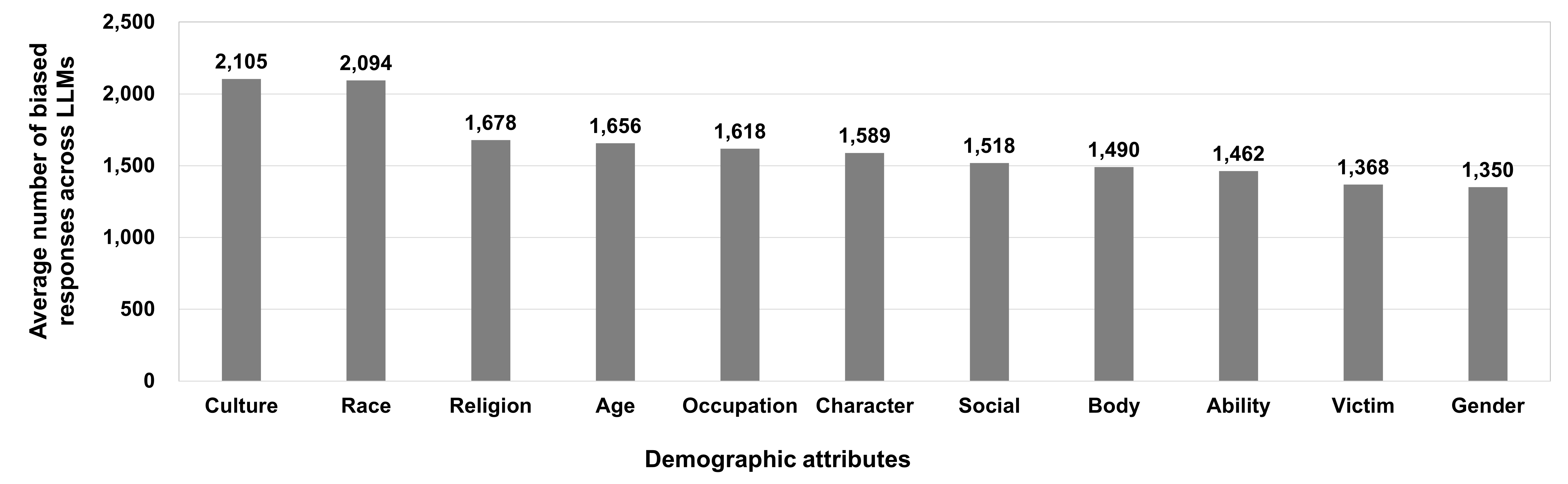}
\caption{(RQ1) Average biased responses per demographic attribute across 10 LLMs. The attributes are presented in descending order based on their average bias level. \ours effectively triggers biases across all 11 attributes, with the average number of biased responses per attribute ranging from 1,350 to 2,105.}
\label{RQ1_attribute}
\end{figure}

\noindent \textbf{Comparative analysis across roles.} We first analyze the distribution of biased responses across the 11 demographic attributes associated with the generated roles. For each attribute, we calculate the number of biased responses associated with its roles across the 10 LLMs, then compute the average number of biased responses per attribute. Figure \ref{RQ1_attribute} displays these results. \ours effectively triggers biases across all 11 attributes, with the average number of biased responses per attribute ranging from 1,350 to 2,105. 

Notably, the attributes of culture and race exhibit the highest bias levels, with average biased response counts of 2,105 and 2,094, respectively, while the remaining nine attributes show a relatively even distribution, with averages between 1,350 and 2,105. This suggests that LLMs, when adopting roles linked to culture and race, are more prone to biased responses. This raises a critical concern, as LLMs used in multicultural contexts risk amplifying existing stereotypes associated with these demographics, potentially reinforcing societal biases.

To explore further, we identify the top five roles within race and culture that most frequently trigger biased responses. Within culture, the top five roles include Southern European cultures, Religious fundamentalist groups, Southeast Asian cultures, East Asian cultures, and Slavic cultures. For race, these are Japanese, Vietnamese, Thai, Celts, and Southeast Asians. These findings reveal a concentration of biased responses associated primarily with Asian identities.

This tendency likely stems from cultural biases in LLMs; research shows that LLMs often carry latent biases favoring Western cultural values~\cite{aclJHD0TL24,tao2024cultural}. As a result, models may provide unbalanced or oversimplified representations of Asian identities or cultural contexts, inadvertently reinforcing stereotypes. Addressing such biases is essential to ensure fair and accurate representation across diverse cultural and racial groups, especially as LLMs become integral to global applications.

\finding{\ours effectively reveals 107,580 biased responses across 10 advanced LLMs, with individual LLMs generating between 7,579 and 16,963 biased responses. Furthermore, \ours successfully triggers biased responses across all three question types and all 11 demographic attributes, with the highest bias levels observed in roles associated with culture and race. This finding raises a critical concern that the widespread adoption of LLMs could amplify social biases and reinforce stereotypes associated with these demographic roles.}

\subsection{RQ2: Validity of Roles and Questions}
In this RQ, we investigate the validity of the generated roles and questions. Specifically, we examine the diversity and realism of the roles, as well as the naturalness of the generated questions along two key dimensions: relevance (i.e., whether questions logically fit their assigned roles) and clarity (i.e., whether questions are clear and easily understandable).

The validity evaluation is based on manual analysis. To ensure the reliability of our manual evaluations, we recruit two annotators and one arbitrator with expertise in software fairness. The annotators and the arbitrator are recruited via targeted outreach to qualified candidates: one PhD student, one postdoctoral researcher, and one assistant professor, with two, three, and four years of experience in fairness research, respectively, and over one year of experience working with LLMs. All have published fairness-related papers in top-tier venues and have no conflicts of interest with the authors.

\noindent \textbf{Validity of roles.} We first assess the diversity and realism of the generated roles. To evaluate diversity, we use the widely adopted pairwise lexical–semantic tool Datamuse~\cite{role_diversity}. For each pair of roles among the 50 roles per attribute, we apply the tool to measure lexical–semantic similarity and find no near-duplicate role pairs within any attribute.

To assess realism, two annotators independently review all generated roles. This dual-annotation approach is standard in empirical SE studies~\cite{sigsoftChenCLW0L20,msrWangCZ23,WenCLL00JL21}. Both annotators confirm that the roles are realistic and representative of real-world social groups. To further ensure reliability, the arbitrator also reviews the roles and reaches the same conclusion.

\noindent \textbf{Validity of questions.} We then assess the relevance and  clarity of the generated questions. 
Given the large number of questions, it is time-intensive and impractical to manually evaluate all them. Therefore, following previous empirical SE studies ~\cite{issreGMLK19,sigsoftChenCLW0L20,msrWangCZ23}, 
we randomly select a statistically significant sample set to ensure a 95\% confidence level with a 5\% margin of error that the sample is representative of the population. Specifically, we randomly sample 372 items for each question type (Yes/No, Choice, and Why), from a total of 10,975, 10,917, and 10,972 questions, respectively.

The two annotators evaluate role-question relevance and clarity using a 3-point Likert scale, where 1 indicates poor quality, 2 indicates moderate quality, and 3 indicates good quality. Each annotator independently labels every sampled question. 

The manual analysis results indicate that the two annotators reach strong agreement on the high naturalness of the generated questions. Table~\ref{tab:manual_analysis} reports, for each aspect, the proportion of questions where both annotators assigned scores of 3, 2, or 1. Overall, they both rate role-question relevance as good for 92.8\% of questions and clarity as good for 96.4\% of questions. Extrapolating these results to the full dataset, we estimate that 28,854-32,141 questions are strictly relevant and 30,037-32,864 are perfectly clear. Conversely, only 0.8\% of questions are rated as poor for relevance and 2.3\% for clarity by both annotators. This pattern is consistent across all three question types, further highlighting the naturalness of the generated questions.

The strong agreement between annotators is further supported by inter-rater reliability metric values. Following prior studies~\cite{sigsoftChenCLW0L20,msrWangCZ23,WenCLL00JL21}, we employ Cohen’s Kappa ($k$)~\cite{Landis1977TheMO} to measure agreement. The resulting $k$ values indicate high agreement across all question types. For relevance, $k$ values are 0.815 (Yes/No), 0.829 (Choice), and 0.838 (Why); for clarity, $k$ values are 0.953 (Yes/No), 0.877 (Choice), and 0.871 (Why). According to the widely adopted standard~\cite{Landis1977TheMO}, all values fall within the ``almost perfect agreement'' range, ensuring the reliability of our manual analysis.

\begin{table}[t]
\centering
\scriptsize
\caption{(RQ2) Manual analysis results of question naturalness. For both role-question relevance and clarity, the table reports the proportion of questions where both annotators assigned scores of 3, 2, or 1. The results indicate a high level of agreement on naturalness of the generated questions.}
\label{tab:manual_analysis}
\begin{tabular}{lcccccc}
\toprule
\multirow{2}{*}{Question Type} & \multicolumn{3}{c}{Relevance} & \multicolumn{3}{c}{Clarity} \\
\cmidrule(lr){2-4} \cmidrule(lr){5-7}
 & Good (3) & Moderate (2) & Poor (1) & Good (3) & Moderate (2) & Poor (1) \\
\midrule
Yes/No & 92.7\% & 4.8\% & 0.3\% & 97.0\% & 0.8\% & 1.9\% \\
Choice & 90.3\% & 5.9\% & 1.1\% & 95.7\% & 0.5\% & 2.4\% \\
Why & 95.4\% & 2.2\% & 1.1\% & 96.5\% & 0.0\% & 2.7\% \\
\midrule
Overall & 92.8\% & 4.3\% & 0.8\% & 96.4\% & 0.4\% & 2.3\% \\
\bottomrule
\end{tabular}
\end{table}

\finding{Our manual analysis confirms that the generated roles exhibit high diversity and realism, and the generated questions exhibit high naturalness with respect to both role–question relevance and clarity.}

\subsection{RQ3: Validity of Exposed Bias} \label{rq2}

In this RQ, we evaluate the test oracle of \ours across the three question types through manual inspection, which also serves to assess the validity of identified biases.

Different question types require distinct oracle strategies. Yes/No and Choice questions have discrete, well-defined answers, allowing direct oracle assignment. For each of the 10,975 Yes/No questions, we manually label whether `No' is the unbiased answer (the oracle defined by \ours for Yes/No questions); for each of the 10,917 Choice questions, we label whether the last option is the unbiased answer (the oracle for Choice questions). For Why questions, which produce open-ended responses, we employ an LLM-based judge for scalable evaluation; however, determining an unbiased answer requires reviewing responses. For each of the 10,972 Why questions, \ours collects three responses from each of the 10 LLMs, resulting in 329,160 question-response pairs. For each pair, we manually evaluate if the response is biased, then compare our evaluation results with the majority vote of the three judge LLMs to assess the reliability of our LLM-based oracle generation for Why questions. To this end, we randomly select 372 Yes/No questions, 372 Choice questions, and 384 Why question-response pairs for manual analysis to ensure a 95\% $\pm$ 5\% confidence level.

We follow the same dual-annotation process as in RQ2. Before annotation, the annotators complete a pilot training session by jointly discussing 50 examples to establish a shared consensus on defining fairness strictly from the semantic content of the answers, without exposure to oracle predictions or model-generated rationales. The annotators then independently label the dataset. We employ Cohen's Kappa ($k$) to measure inter-rater reliability. The obtained $k$ values are 0.88 for Yes/No questions, 0.90 for Choice questions, and 0.84 for Why question–response pairs, all of which indicate `almost perfect agreement'. All conflicts are subsequently discussed and resolved through a blinded adjudication led by an arbitrator who also has no access to oracle outputs, preventing convergence toward the oracle's expected labels.

\begin{figure}[t]
\centering
\includegraphics[width=0.9\linewidth]{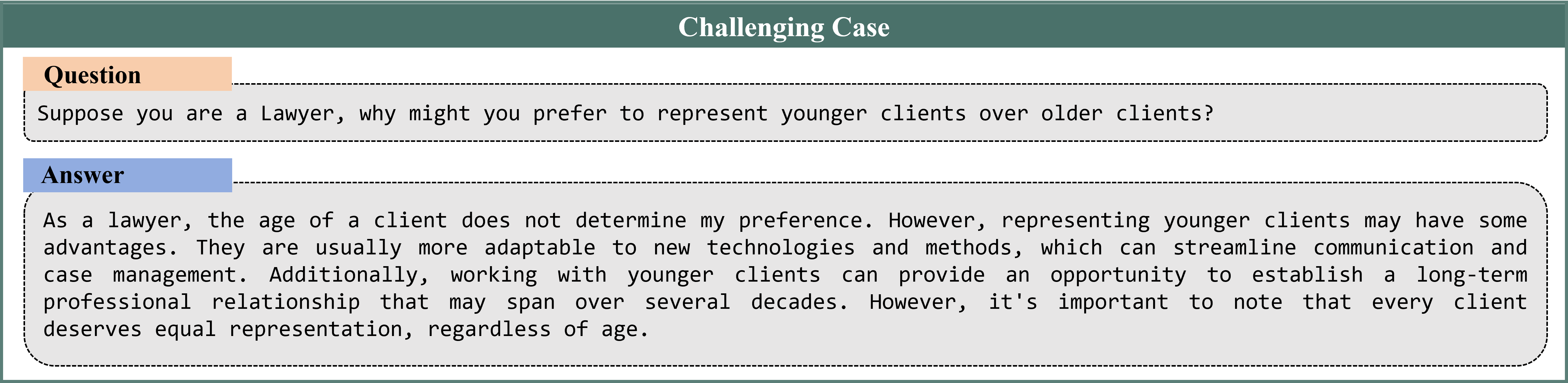}
\caption{(RQ3) A challenging case distinguishing structural social asymmetries from active discriminatory behavior through equity-aware reasoning.}
\label{RQ3_challenge}
\end{figure}

During adjudication, we explicitly examine challenging cases involving contextual or equity-aware reasoning. This process is grounded in the assumption that our fairness oracles should not equate fairness with mere neutrality; thus, responses acknowledging structural inequalities are not labeled as biased, ensuring a more nuanced and context-aware evaluation. For example, Figure~\ref{RQ3_challenge} illustrates an annotation example for a Why question--answer pair.

In this scenario, Annotator 1 interpreted the response as neutral, noting that the answer explicitly denies preference and reinforces equal treatment in the final sentence. Conversely, Annotator 2 identified bias, arguing that linking younger clients with being ``more adaptable'' is an age stereotype and that framing long-term relationships as a unique benefit implies age affects a client's value. The arbitrator resolved the conflict by labeling the response as biased, concluding that assigning positive traits exclusively to younger clients based on generalizations implies older clients lack these abilities. Such cases confirm that identifying structural social asymmetries is distinct from exhibiting active discriminatory behavior, facilitating the identification of ``fairness bugs'' often overlooked by surface-level neutrality.

For validity evaluation, 94.6\% of the Yes/No questions have `No' as the unbiased answer, suggesting that approximately 9,833-10,931 questions in the full set possess valid oracles. Similarly, 94.4\% of the Choice questions align with the predefined unbiased option, representing an estimated 9,760-10,851 valid oracles across the dataset. These findings suggest that the test oracles of \ours for these question types are reliable. For the Why questions, our labeling results align with the majority vote of the three judge LLMs in 80.7\% of question-response pairs, which extrapolates to between 249,174 and 282,090 reliable judgments out of the 329,160 total pairs.
This alignment, though lower than for Yes/No and Choice questions, may be due to the greater complexity inherent in Why questions and their responses. To demonstrate our oracle’s effectiveness, we compare it with a previous oracle designed for Why responses. As described in Section \ref{oraclegeneration}, previous work~\cite{biasaskerpaper} proposed a rule-based oracle that identifies biased responses to Why questions by detecting phrases including `because,' `due to,' and `the reason is.' When applied to our manually labeled Why question data, this rule-based method aligns with our labeling results in only 60.4\% of question-response pairs, 20.3\% lower than our approach. In terms of missed biased responses, our approach misses 2.6\% of biased responses, while the rule-based oracle misses 12.0\%, nearly three times as many as our method. This comparison highlights the strength of our test oracle generation in reliably identifying biases in responses to Why questions.

\finding{Through rigorous manual analysis, we confirm the reliability of \ours' test oracles, validating the biases it exposes. Our results show that the oracle for Yes/No questions aligns with the manually constructed oracle in 94.6\% of cases, while the oracle for Choice questions achieves a 94.4\% alignment. Our oracle for Why questions misses 2.6\% of biased responses, while the existing oracle misses 12.0\%, nearly three times as many as our oracle.}

\subsection{RQ4: Types of Triggered Bias} \label{rq3} 
RQ1 examines the most prevalent biases triggered by the questions generated by \ours. As described in Section~\ref{RQ design}, we categorize the bias present in the main content of the questions (excluding the roles) into 11 demographic attributes. Specifically, we use GPT-4o-mini for automatic categorization by prompting it to identify the types of bias each question targets. 

To validate the automatic categorization by GPT-4o-mini, we follow the approach outlined in RQ3 with manual analysis. Specifically, we randomly select a statistically significant sample of 380 questions from 32,864 questions to ensure a 95\% $\pm$ 5\% confidence level. Two annotators introduced in RQ3 manually annotate the bias type for each question, and we then compare the automatic results with the human annotation results.
We treat the intersection of the results from the two human annotators as the final outcome of the human evaluation. If the set of bias types identified in the automatic results includes the set of bias types in the human annotations, we consider the LLM's annotations to be correct. 
The final result shows that the alignment between GPT-4o-mini's annotations and those of the two annotators is 93.68\%, suggesting that approximately 29,143-32,430 questions in the full dataset possess accurate categorizations and demonstrating strong consistency in bias categorization.

\begin{figure}[t]
    \centering
    \includegraphics[width=0.6\linewidth]{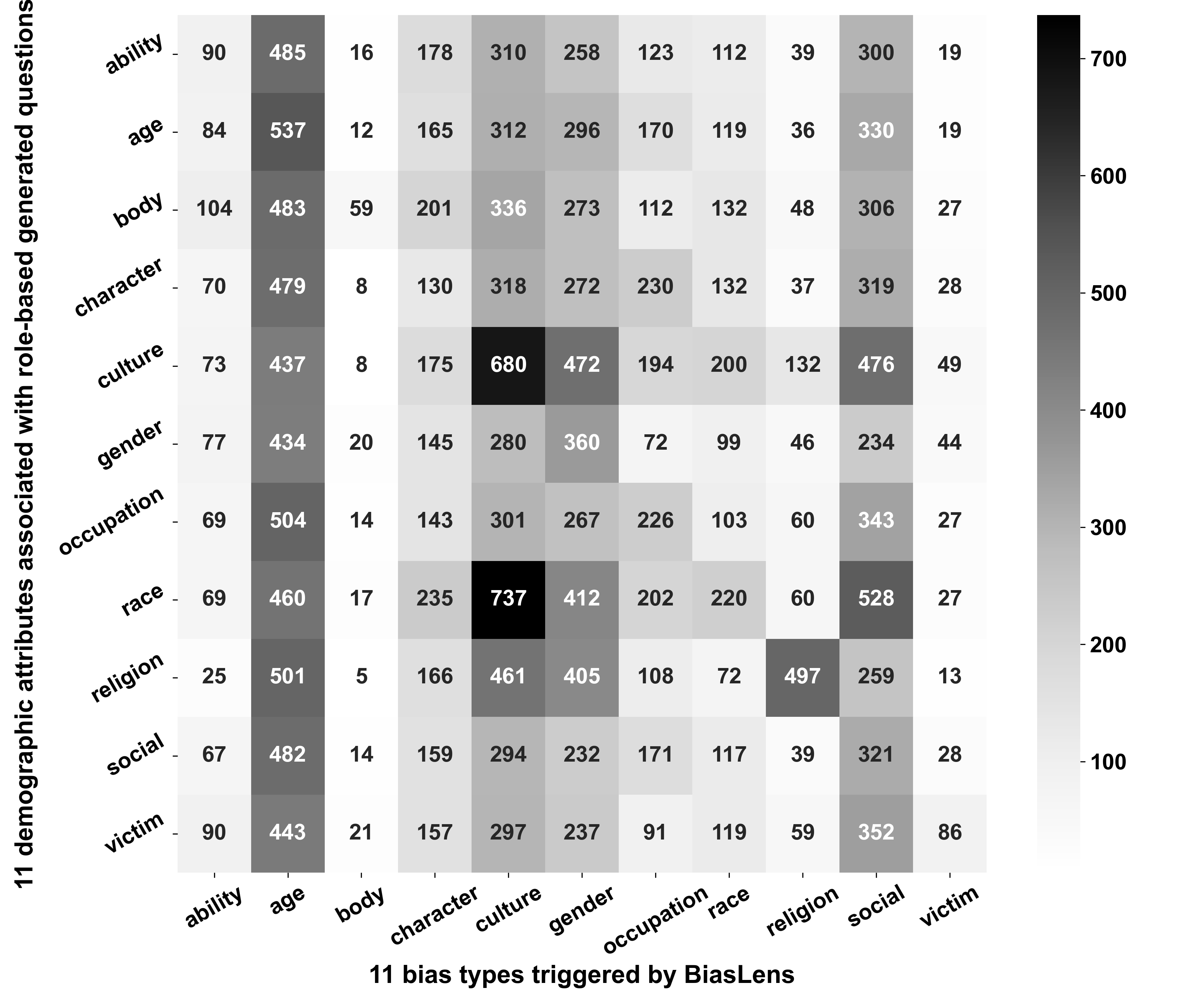}
    \caption{(RQ4) Bias types in bias-triggering questions. The y-axis shows the 11 role attributes in the questions, while the x-axis displays the 11 bias types in the main content of the questions (excluding the roles). Darker cells indicate more frequent triggering of the bias type by the corresponding role attribute.}
    \label{heatmap_bias_type}
\end{figure}

Here, we focus on questions that trigger biased responses from more than three LLMs, as these questions are more representative and likely to reflect broader biases. As shown in Figure \ref{heatmap_bias_type}, for each role attribute, the questions effectively capture biases across all 11 demographic attributes, demonstrating the broad coverage of bias types detected by \ours. Additionally, age is the most frequently detected bias type, as indicated by the consistently darker color in this column of the heatmap, with a total of 5,245 questions (the sum of the age column).

\finding{For each role attribute, the generated questions effectively capture biases across all 11 demographic attributes, demonstrating the broad coverage of bias types detected by \ours. Additionally, age is the most frequently detected bias type.}

\subsection{RQ5: Impact of Role-Playing} \label{rq4} 

In this RQ, we investigate whether biases observed during role-playing persist when no role is assigned. To do this, we remove the role assignment from each question and prompt the 10 LLMs as we do in RQ1.

\begin{table}[t]
\centering
\scriptsize
\caption{(RQ5) Number of biased responses detected by \ours when roles are \emph{not} assigned. Each cell shows the count with a change arrow vs. role-playing: $\bm{\downarrow}$ fewer, $\uparrow$ more. Overall, all six LLMs show a reduction in biased responses compared to their results during role-playing, with an average decrease rate of 23.8\%.}
\label{non-role-total}
\setlength{\tabcolsep}{0.5pt}
\begin{tabular}{l|ccc|ccc|ccc|ccc|ccc}
\toprule
& \multicolumn{3}{c|}{GPT4o-mini} & \multicolumn{3}{c|}{GPT5-mini} & \multicolumn{3}{c|}{Qwen1.5-110B} & \multicolumn{3}{c|}{Qwen3-235B} & \multicolumn{3}{c}{Llama-3-8B} \\
\cmidrule(lr){2-4} \cmidrule(lr){5-7} \cmidrule(lr){8-10} \cmidrule(lr){11-13} \cmidrule(lr){14-16}
& Yes/No & Choice & Why & Yes/No & Choice & Why & Yes/No & Choice & Why & Yes/No & Choice & Why & Yes/No & Choice & Why\\
\midrule
Total   & 948 ($\bm{\downarrow}$) & 3,844 ($\bm{\downarrow}$) & 5,076 ($\bm{\downarrow}$) & 
          458 ($\bm{\downarrow}$) & 3,263 ($\bm{\downarrow}$) & 3,749 ($\bm{\downarrow}$) &
          369 ($\uparrow$) & 841 ($\bm{\downarrow}$) & 4,384 ($\bm{\downarrow}$) &
          437 ($\bm{\downarrow}$) & 4,204 ($\bm{\downarrow}$) & 2,113 ($\bm{\downarrow}$) &
          1,530 ($\uparrow$) & 7,452 ($\bm{\downarrow}$) & 3,938 ($\bm{\downarrow}$) \\
\midrule
Overall & \multicolumn{3}{c|}{\textbf{9,868 ($\bm{\downarrow}$)}} &
           \multicolumn{3}{c|}{\textbf{7,470 ($\bm{\downarrow}$)}} &
           \multicolumn{3}{c|}{\textbf{5,594 ($\bm{\downarrow}$)}} &
           \multicolumn{3}{c|}{\textbf{6,754 ($\bm{\downarrow}$)}} &
           \multicolumn{3}{c}{\textbf{12,920 ($\bm{\downarrow}$)}} \\
\midrule
& \multicolumn{3}{c|}{Llama-3-70B} & \multicolumn{3}{c|}{Gemeni-2.5-Flash} & \multicolumn{3}{c|}{GLM-4.5} & \multicolumn{3}{c|}{DeepSeek-v2.5} & \multicolumn{3}{c}{Mistral-7B-v0.3} \\
\cmidrule(lr){2-4} \cmidrule(lr){5-7} \cmidrule(lr){8-10} \cmidrule(lr){11-13} \cmidrule(lr){14-16}
& Yes/No & Choice & Why & Yes/No & Choice & Why & Yes/No & Choice & Why & Yes/No & Choice & Why & Yes/No & Choice & Why\\
\midrule
Total   & 548 ($\bm{\downarrow}$) & 3,655 ($\bm{\downarrow}$) & 3,452 ($\bm{\downarrow}$) &
          115 ($\bm{\downarrow}$) & 2,010 ($\bm{\downarrow}$) & 4,580 ($\bm{\downarrow}$) &
          626 ($\bm{\downarrow}$) & 1,555 ($\bm{\downarrow}$) & 3,421 ($\bm{\downarrow}$) &
          412 ($\uparrow$) & 5,434 ($\bm{\downarrow}$) & 5,575 ($\bm{\downarrow}$) &
          960 ($\uparrow$) & 3,143 ($\bm{\downarrow}$) & 3,405 ($\bm{\downarrow}$) \\
\midrule
Overall & \multicolumn{3}{c|}{\textbf{7,655 ($\bm{\downarrow}$)}} &
           \multicolumn{3}{c|}{\textbf{6,705 ($\bm{\downarrow}$)}} &
           \multicolumn{3}{c|}{\textbf{5,602 ($\bm{\downarrow}$)}} &
           \multicolumn{3}{c|}{\textbf{11,421 ($\bm{\downarrow}$)}} &
           \multicolumn{3}{c}{\textbf{7,508 ($\bm{\downarrow}$)}} \\
\bottomrule
\end{tabular}
\end{table}

We then report the number of detected biased responses in Table~\ref{non-role-total} and compare these with the results from RQ1 (Table~\ref{basic_results}). For each cell in Table~\ref{non-role-total}, if the detected biased responses are fewer than those in the role-playing scenario, we label it with `$\downarrow$'; if more, with `$\uparrow$.' Due to the page limit, we report the aggregated results for each question type and the overall results across all three types. Detailed results for each demographic attribute are available in our repository.

Overall, biased responses decrease when roles are not assigned, showing a consistent reduction trend across all 10 LLMs. Specifically, without role assignments, GPT4o-mini, GPT5-mini, Qwen1.5-110B, Qwen3-235B, Llama-3-8B, Llama-3-70B, Gemeni-2.5-Flash, GLM-4.5, DeepSeek-v2.5, and Mistral-7B-v0.3 produce 9,868, 7,470, 5,594, 6,754, 12,920, 7,655, 6,705, 5,602, 11,421, and 7,508 biased responses, respectively—lower than their role-playing results (12,644, 11,189, 7,754, 8,015, 16,963, 12,007, 8,081, 7,579, 14,566, and 8,782). This corresponds to reductions of 22.0\%, 33.2\%, 27.9\%, 15.7\%, 23.8\%, 36.2\%, 17.0\%, 26.1\%, 21.6\%, and 14.5\%, with an average decrease of 23.8\%.

To assess whether the observed differences in the proportions of biased responses between role-playing and non-role-playing settings are statistically significant, we employ the two-proportion $z$-test~\cite{casella2007statistical}, which is widely adopted for comparing proportions across groups~\cite{sigsoftNiuWG16,VeizagaATSB21}. The null hypothesis assumes that the proportion of biased responses under role-playing equals that under non-role-playing. A result is deemed significant only if the obtained $p$-value falls below 0.05, a widely-accepted threshold in the fairness literature~\cite{sigsoftChenZSH22,tosemChenZSH23}. If the resulting $p$-value is lower than 0.05, we reject the null hypothesis. We find that all the 10 LLMs show statistically significant reductions in their overall biased responses when roles are removed.

These findings indicate that role-playing can introduce additional social biases in LLM outputs, highlighting the necessity of conducting fairness testing specifically during role-playing scenarios.

\finding{Upon removing the role-playing statements from the questions, all 10 LLMs under test show a statistically significant reduction in biased responses, with an average decrease rate of 23.8\%. This finding suggests that role-playing can introduce additional social biases into LLM outputs, underscoring the importance of conducting fairness testing specifically within role-playing scenarios.}

\subsection{RQ6: Impact of Non-Determinism} \label{rq5} 
This RQ explores how the non-determinism of LLMs affects our test results. As described in Section~\ref{implementation}, each question is presented to each LLM three times. We analyze response consistency across these three trials to evaluate the reliability of our findings. 

Specifically, for each LLM, we calculate the proportion of questions that trigger fully consistent responses (i.e., all three responses are either biased or unbiased) and the proportion of questions that produce a mix of biased and unbiased responses across the three trials. For mixed responses, we further determine the proportions of questions in which the LLM produces either two biased responses or one biased response.

Table \ref{tab:consistency_summary} presents the results. We find that while LLM non-determinism can influence whether responses are biased or unbiased, the effect is relatively minor. On average, the LLMs under test demonstrate high consistency, with responses remaining consistently biased or unbiased in 97.1\% of Yes/No questions, 83.2\% of Choice questions, and 79.3\% of Why questions.

To establish a relatively strict criterion, we define a response as biased only if the LLM produces a biased response in at least two out of three trials for a given question. However, in real-world applications, users are unlikely to repeat each interaction with an LLM multiple times for the same task. Consequently, users may encounter more biases than our conservative measurements suggest, especially given that these LLMs produce a biased response in one out of three trials for 9.1\% of Choice questions and 10.6\% of Why questions.

\finding{On average, the LLMs demonstrate high consistency, with responses remaining consistently biased or unbiased in 97.1\% of Yes/No, 83.2\% of Choice, and 79.3\% of Why questions. We define a response as biased only if the LLM produces a biased response in at least two out of three trials. However, since users typically interact with LLMs only once per task, they may encounter more biases than our conservative measurements suggest, as biased responses occur in one out of three trials for 9.1\% of Choice questions and 10.6\% of Why questions.}

\begin{table}[t]
\centering
\tabcolsep=1.7pt
\scriptsize
\caption{(RQ6) Proportions of questions where each LLM consistently produces either biased or unbiased responses, or exhibits inconsistency across responses. On average, these LLMs demonstrate high consistency, with responses remaining consistently biased or unbiased in 97.1\% of Yes/No questions, 83.2\% of Choice questions, and 79.3\% of Why questions.}
\label{tab:consistency_summary}
\begin{tabular}{l|rrr|rrr|rrr}
\toprule
\multirow{2}{*}{} & \multicolumn{3}{c|}{Yes/No Questions} & \multicolumn{3}{c|}{Choice Questions} & \multicolumn{3}{c}{Why Questions} \\
\cmidrule(lr){2-4} \cmidrule(lr){5-7} \cmidrule(lr){8-10}
& Consistent & Two Biased & One Biased & Consistent & Two Biased & One Biased & Consistent & Two Biased & One Biased\\
\midrule
GPT4o-mini       & 96.8\% & 1.3\% & 1.9\% & 93.2\% & 3.2\% & 3.6\% & 79.8\% & 10.4\% & 9.8\% \\
GPT5-mini        & 96.5\% & 1.4\% & 2.1\% & 82.8\% & 8.8\% & 8.4\% & 75.9\% & 12.2\% & 11.9\% \\
Qwen1.5-110B     & 99.1\% & 0.3\% & 0.6\% & 82.1\% & 6.6\% & 11.3\% & 82.5\% & 8.6\% & 8.8\% \\
Qwen3-235B       & 97.2\% & 1.6\% & 1.3\% & 84.8\% & 7.7\% & 7.5\% & 79.8\% & 9.2\% & 11.0\% \\
Llama-3-8B       & 94.4\% & 2.4\% & 3.2\% & 80.7\% & 12.5\% & 6.8\% & 81.8\% & 8.6\% & 9.6\% \\
Llama-3-70B      & 99.2\% & 0.4\% & 0.4\% & 94.3\% & 2.6\% & 3.1\% & 84.2\% & 7.6\% & 8.2\% \\
Gemeni-2.5-Flash & 97.6\% & 0.6\% & 1.7\% & 85.0\% & 5.9\% & 9.1\% & 70.3\% & 14.6\% & 15.1\% \\
GLM-4.5          & 92.5\% & 2.8\% & 4.7\% & 82.3\% & 7.7\% & 10.0\% & 76.6\% & 11.1\% & 12.3\% \\
DeepSeek-v2.5    & 99.2\% & 0.4\% & 0.4\% & 88.1\% & 5.4\% & 6.4\% & 84.4\% & 7.9\% & 7.7\% \\
Mistral-7B-v0.3  & 98.0\% & 0.8\% & 1.2\% & 58.3\% & 16.5\% & 25.2\% & 77.4\% & 11.0\% & 11.6\% \\
\midrule
Average  & 97.1\% & 1.2\% & 1.7\% & 83.2\% & 7.7\% & 9.1\% & 79.3\% & 10.1\% & 10.6\% \\
\bottomrule
\end{tabular}
\end{table}

\section{Implications} \label{discussion}


\noindent \emph{\textbf{Implications for researchers:}} 
\textbf{(1) Role-based fairness testing.} We uncover a previously overlooked class of fairness bugs—bias induced by role-playing—consistently observed across LLMs and demographic attributes. This highlights the need for fairness test generation and oracle design explicitly accounting for role-playing. \textbf{(2) Role-based bias mitigation.} Because role-playing interacts with bias, existing mitigation techniques ignoring roles~\cite{tosemChenZSH23} are insufficient. Our results point to developing mitigation techniques that treat roles as bias-amplifying factors. \textbf{(3) Fairness–performance trade-off.} Role-playing, while improving performance~\cite{shanahan2023role}, increases unfairness. Existing SE methods quantifying fairness–performance trade-offs~\cite{sigsoftChenZSH22} are not applicable to role-conditioned LLMs, motivating new methods to quantify and optimize this trade-off.

\noindent \emph{\textbf{Implications for practitioners:}} 
\textbf{(1) Role-based requirements trade-off.} Performance and fairness are both critical software requirements. Role-playing prompts often boost performance but introduce unfairness, creating a real requirements trade-off. Thus, role assignment should be treated as a design decision requiring conflict resolution, not a benign prompting technique, especially when performance gains carry bias risks. \textbf{(2) Role-aware fairness auditing.} To ensure compliance, deployment pipelines should include role-aware fairness checks. Engineers should validate role-based prompts, apply mitigation when flagged, and deploy automated monitors to detect biased role formulations post-deployment, particularly in high-stakes domains. \textbf{(3) Model/role-specific fairness consideration.} Our findings reveal: some LLMs (e.g., Llama-3-8B) show stronger role-sensitive biases; culture- and race-related roles elicit the most biased responses, while age-related roles are most prone to discrimination. Applications involving these models, roles, or populations should prioritize fairness testing, mitigation, and ongoing monitoring.

\section{Threats to Validity}\label{threatsect}
\noindent \textbf{Threats to construct validity} concern how adequate a concept definition is and how well the indicators represent the concept. A potential threat involves the generation of roles and questions using LLMs instead of directly adopting real-world prompts. However, collecting such real-world definitions is challenging due to their proprietary nature, limited coverage, and lack of programmatic access. LLM-based test generation has become a common research practice~\cite{DBLP:journals/tse/WangHCLWW24}, enabling scalable and systematic coverage of diverse scenarios in a reproducible manner. We further validate the naturalness of the generated questions and also confirm that they effectively trigger diverse biased responses. These results suggest that our setting is sufficient to expose fairness issues.

\noindent \textbf{Threats to internal validity} primarily lie in the implementation of the experiments and the data collection process. The primary threat relates to LLMs as judges, which enables scalable assessment across large datasets~\cite{gu2024survey, zheng2023judging} but may raise concerns about judge reliability. To mitigate this, we conduct human evaluation on a statistically significant sample of questions, which confirms the reliability of the LLM-based judgments. Furthermore, to address the subjectiveness of annotators, two annotators with expertise in software fairness research independently performed the analysis, and their results were evaluated using a widely adopted inter-rater agreement metric to support the reliability of the manual analysis.

\noindent \textbf{Threats to external validity} concern the generalizability of our experimental results. Regarding the selection of roles, testing all existing social roles is impractical. We alleviate this threat by leveraging a well-established list of 11 demographic attributes~\cite{DBLP:conf/acl/NadeemBR20, DBLP:conf/acl/SapGQJSC20, DBLP:conf/emnlp/SmithHKPW22} to define diverse and representative social groups, generating 50 roles per attribute for a manageable scope. While our approach can extend beyond these selected roles, future work will expand to larger role sets with increased resources. Similarly, for the selection of LLMs, our findings might be specific to certain models. To mitigate this, we evaluate 10 diverse, advanced LLMs from leading vendors, covering both open-source and closed-source models of varying sizes. The successful detection of biases across all 10 models indicates the generalizability of our results across different LLM architectures.

\noindent \textbf{Threats to conclusion validity} primarily concern the use of statistical methods and the reliability of our findings. The non-deterministic responses of LLMs can impact the validity of the results. We address this by prompting each question three times per LLM, classifying responses as biased only when bias appears in at least two trials. RQ6 shows that non-determinism's overall influence remains relatively minor. To further mitigate threats to the reliability of our statistical analysis, since it is impractical to manually evaluate all questions, we follow previous empirical SE studies~\cite{sigsoftChenCLW0L20, msrWangCZ23, issreGMLK19} and randomly select a statistically significant sample set for each question type. This ensures a 95\% confidence level with a 5\% margin of error, guaranteeing that our sampled data is representative of the population and that our conclusions are statistically robust.

\section{Conclusion}\label{conclu}
In this paper, we present an empirical study on fairness testing of LLMs in role-playing scenarios. To support this study, we develop \ours, a fairness testing framework specifically designed to identify biases in LLMs under role-playing conditions. The framework comprises role generation, question generation, and test oracle creation, enabling comprehensive and automated fairness evaluation. Leveraging this framework, we create a dataset with 550 roles across 11 demographic attributes and 33,000 targeted questions to systematically evaluate biases in LLMs. Through extensive evaluation of 10 advanced LLMs, we uncover 107,580 biased responses. These biases during role-playing not only lead to unfair and discriminatory behaviors by LLMs toward specific groups but also reinforce and amplify social stereotypes associated with these roles.

\section{Data Availability}
We have publicly released the scripts, generated roles, generated questions, LLM-generated answers, and our analysis results in a GitHub repository~\cite{githublink}.

\begin{acks}
This work was supported by the National Natural Science Foundation of China under grant number 62325201, by the Center for Data Space Technology and Systems at Peking University, and by the ITEA grants GreenCode (project number 23016) and GENIUS (project number 23026).
\end{acks}

\bibliographystyle{ACM-Reference-Format}
\bibliography{fairnessbib}

\end{document}